\documentclass[12pt,reqno]{amsart}

\usepackage{epsfig}
\usepackage{amsmath, amsthm, amsfonts}
\usepackage{graphicx}

\numberwithin{equation}{section}

\newcommand{\R}{{\mathbb R}}

\newcommand{\Z}{{\mathbb Z}}

\newcommand{\be}{\beta}
\newcommand{\ga}{\gamma}

\newcommand{\la}{\lambda}
\newcommand{\ep}{\varepsilon}
\newcommand{\de}{\delta}

\newcommand{\f}{\varphi}
\newcommand{\sg}{\sigma}
\newcommand{\Sg}{\Sigma}

\newtheorem{theo}{{\sc \bf Theorem}}[section]

\newtheorem{prop}[theo]{{\sc \bf Proposition}}

\begin{document}

\title[Hierarchical models and renornalization group]
{Critical phenomena in the Dyson hierarchical model and renormalization group}

\author{Pavel Bleher}
\address{Department of Mathematical Sciences,
Indiana University-Purdue University Indianapolis,
402 N. Blackford St., Indianapolis, IN 46202, U.S.A.}
\email{bleher@math.iupui.edu}

\thanks{The author is supported in part
by the National Science Foundation (NSF) Grants DMS-0652005 and DMS-0969254.}

\date{\today}

\begin{abstract} We review some results on the critical phenomena in the Dyson hierarchical model and renormalization group.
\end{abstract}

\maketitle

\section{The Dyson hierarchical model}

The Dyson hierarchical model was introduced by Dyson \cite{Dys1}--\cite{Dys3}. It is 
defined as follows. Consider the set 
\begin{equation}\label{v1}
V_n=\{1,2,\ldots,2^n\},
\end{equation}
consisting of $2^n$ elements, and for any $j=0,1,\ldots,n$, the partition $\pi_j$ of $V_n$, 
\begin{equation}\label{v2}
V_n=\bigsqcup_{k=1}^{2^{n-j}} V_{jk},
\end{equation}
into the subsets
\begin{equation}\label{v3}
V_{jk}=\{(k-1)2^j+1,(k-1)2^j+2,\ldots,k 2^j\}.
\end{equation}
Each set $V_{jk}$ has $2^j$ elements. The partitions $\pi_j$ are ordered,
\begin{equation}\label{v4}
\pi_0\succ\pi_1\succ\ldots\succ\pi_n.
\end{equation}
In fact, any set $V_{jk}$ of the partition $\pi_j$ consists of the two sets, $V_{j-1,2k-1}$
and $V_{j-1,2k}$, of the partition $\pi_{j-1}$. The partition $\pi_0$ is a partition of $V_n$
into one-point sets, while $\pi_n$ consists of one set, $V_{n1}=V_n$.

For any two different points $x,y$ in $V_n$, define the number 
\begin{equation}\label{v5}
j(x,y)=\min\,\{ j:\; \{x,y\}\subset V_{jk}\quad \textrm{for some}\quad k=1,\ldots,2^{n-j}\},
\end{equation}
which is the smallest $j$ such that both $x$ and $y$ lie in one element of the partition $\pi_j$.
Define then the {\it hierarchical distance} between two points $x,y$ in $V_n$ as 
\begin{equation}\label{v6}
d(x,y)=2^{j(x,y)-1}, \quad \textrm{if}\quad x\not=y;\qquad d(x,x)=0.
\end{equation}

A configuration $\sg$ of the Dyson hierarchical model in the set $V_n$ consists of $2^n$
spin variables,
\begin{equation}\label{v8}
\sg=\{\sigma(x)\in\R,\quad x\in V_n\},
\end{equation}
and, respectively, the configuration space of the model in $V_n$ is 
\begin{equation}\label{v9}
\Sg_n=\R^{(2^n)}.
\end{equation}
The Hamiltonian of the  Dyson hierarchical model is defined as
\begin{equation}\label{v10}
H_n(\sg)=-J\sum_{\{x,y\}\subset V_n,\;x\not=y}\frac{\sg(x)\sg(y)}{d(x,y)^a}\,,
\end{equation}
where $a>0$ is a parameter of the Dyson hierarchical model. 
The number $J$ is the interaction constant, and $J>0$ corresponds to the ferromagnetic
model. In what follows we will assume that $J>0$. Moreover, for the sake of simplicity, we will 
assume that $J=1$.

Let $\nu$ be a probability measure on $\R$ such that 
\begin{equation}\label{v11}
\int_{\R} e^{As^2}d\nu(s)<\infty,\qquad \forall A>0.
\end{equation}
The Gibbs distribution $\mu_n$ of the Dyson hierarchical model in the set $V_n$ is defined as
\begin{equation}\label{v12}
d\mu_n(\sg)=\frac{1}{Z_n}\,e^{-\be H_n(\sg)}\prod_{x\in V_n} d\nu(\sg(x)),
\end{equation}
where 
\begin{equation}\label{v13}
\be=\frac{1}{T}\ge 0
\end{equation}
is the reciprocal temperature and
\begin{equation}\label{v14}
Z_n=\int_{\Sg_n} e^{-\be H_n(\sg)}\prod_{x\in V_n} d\nu(\sg(x))
\end{equation}
is the partition function. When $\be=0$, formula (\ref{v12}) reduces to
\begin{equation}\label{v15}
d\mu_n(\sg)=\prod_{x\in V_n} d\nu(\sg(x)),
\end{equation}
so that in this case $\sg(x),\; x\in V_n,$ are independent identically distributed 
 random variables with the distribution $\nu$.

\section{Renormalization group transformation of the Dyson hierarchical model}

Let $\mu$ be a probability distribution on the configuration space $\Sg_n=\R^{(2^n)}$.
The renormalization group (RG) transformation of $\mu$ is defined as follows. For any $x=1,\ldots,2^{n-1}$,
the set $V_{1x}$ of the partition $\pi_1$ consists of two elements,
\begin{equation}\label{rg1}
V_{1x}=\{2x-1,2x\}.
\end{equation}
Consider the random variables
\begin{equation}\label{rg2}
\sg'(x)=\frac{\sg(2x-1)+\sg(2x)}{2^{\kappa}},\quad x=1,\ldots,2^{n-1},
\end{equation}
with respect to the probability distribution $\mu$, where $\kappa>0$ is a parameter of the RG
transformation. The random variables $\sg'(x)$,
$x\in V_{n-1}=\{1,\ldots,2^{n-1}\}$, are called
the {\it normalized block spin variables}, with the normalization parameter $\kappa$.
Consider the configuration 
\begin{equation}\label{rg3}
\sg'=\left\{\sg'(x)=\frac{\sg(2x-1)+\sg(2x)}{2^{\kappa}}\,,\quad x\in V_{n-1}\right\}\in \Sg_{n-1}
\end{equation}
of the normalized block spin variables, and its probability distribution, $\mu(\sg')$, with respect
to the Gibbs distribution $\mu$. The probability distribution $\mu(\sg')$ on the
configuration space $\Sg_{n-1}$is called the 
{\it renormalization group transformation} of the probability distribution $\mu$, and it is denoted $R\mu$.

If we perform $m$ iterations of the RG transformation, this produces a probability distribution $R^m\mu$
of the block spin configurations, 
\begin{equation}\label{rg4}
\sg'=\left\{\sg'(x)=2^{-m\kappa}\sum_{y\in V_{mx}}\sg(y)\,,\quad x\in V_{n-m}\right\}\in \Sg_{n-m}.
\end{equation}
In particular, when $m=n$, we get a probability distribution of the normalized sum of spins
$\sg(x)$ in $V_n$,
\begin{equation}\label{rg5}
\sg'=2^{-n\kappa}\sum_{y\in V_{n}}\sg(y)\,.
\end{equation}
Thus the notion of the RG transformation is a generalization of a normalized sum of random variables, and
it is convenient in the study of limit probability laws of random fields. 
In this setting, the value $\kappa=1$
corresponds to the law of large numbers, while $\kappa=\frac{1}{2}$ corresponds to the central limit theorem.
The RG transformation can be extended to random fields on the set $(V_n)^d$ in the $d$-dimensional 
lattice $\Z^d$, if we define it as the distribution of the block spins,  
\begin{equation}\label{rg6}
\sg'(x)=2^{-d\kappa}\sum_{y:\;\{y_j\in V_{1x_j},\;1\le j\le d\}}\sg(y),\quad x\in(V_{n-1})^d,
\end{equation}
on $\Z^d$.

The basic idea of the method of renormalization group in the theory of critical phenomena 
of statistical mechanics \cite{WK} is that at the critical point, there should exist a nontrivial limit 
of the iterated RG transformations $R^m$ of the Gibbs distribution $\mu_n$, with some parameter $\frac{1}{2}<\kappa<1$, 
as simultaneously $m\to\infty$ and $(n-m)\to\infty$.
The limiting random field, 
\begin{equation}\label{rg7}
\mu^*=\lim_{m,\,(n-m)\to\infty} R^m\mu_n,
\end{equation}
is a fixed point of the RG transformation, and the critical exponents can be expressed in terms
of the parameter $\kappa$ and the eigenvalues of the linearized RG transformation at the fixed
point. Mathematically, the development of the RG approach to the critical phenomena is a difficult,
challenging problem, and it was first successfully applied to the Dyson hierarchical model
in the papers of Bleher and Sinai \cite{BS1}, \cite{BS2}.

For  the Dyson hierarchical model, the
RG transformation can be reduced to a nonlinear transformation of the underlying
probability measure $\nu$ on $\R$. Namely, the following proposition holds: 

\begin{prop} \label {hmt}(see \cite{BS1}). If $\mu_n$ is a Gibbs distribution of the Dyson hierarchical model
in the set $V_n$ and $R$ is the RG transformation with the parameter $\kappa=\frac{a}{2}$,  then $R\mu_n$ coincides with
 the Gibbs distribution  of the Dyson hierarchical model
in the set $V_{n-1}$ with respect to the probability measure $\nu'$ on $\R$ such that
\begin{equation}\label{rg8}
\nu'(A)=\frac{1}{Z}\iint_{\frac{s+t}{2^{a/2}}\in A} e^{\be (s,t)} d\nu(s)d\nu(t)
\end{equation}
for any measurable set $A$, where
\begin{equation}\label{rg9}
Z=\iint_{\R\times \R} e^{\be (s,t)} d\nu(s)d\nu(t).
\end{equation}
\end{prop}

\begin{proof} If $x\in V_{1u}$ and $y\in V_{1w}$, where  $u,w\in V_{n-1}$
and $u\not=w$, then
\begin{equation}\label{rg10}
d(x,y)=2d(u,w),
\end{equation}
hence 
\begin{equation}\label{rg11}
\begin{aligned}
H_n(\sg)&=-\sum_{\{x,y\}\subset V_n,\;x\not=y}\frac{(\sg(x),\sg(y))}{d(x,y)^a}\\
&=-\sum_{\{x,y\}\subset V_n,\;d(x,y)=1}\frac{(\sg(x),\sg(y))}{d(x,y)^a}
-\sum_{\{x,y\}\subset V_n,\;d(x,y)\ge 2}\frac{(\sg(x),\sg(y))}{d(x,y)^a}\\
&=-\sum_{u\subset V_{n-1}}(\sg(2u-1),\sg(2u))
-\sum_{\{u,w\}\subset V_{n-1},\;u\not=w}\frac{(\sg'(u),\sg'(w))}{d(u,w)^a}\\
&=-\sum_{u\subset V_{n-1}}(\sg(2u-1),\sg(2u))+H_{n-1}(\sg')\,,
\end{aligned}
\end{equation}
and (\ref{rg8}) follows.
\end{proof}

Proposition \ref{hmt} shows that the RG transformation of the Gibbs
distribution of the Dyson hierarchical model reduces to the transformation of the measure $\nu$,
as described by equation (\ref{rg8}). Therefore, the transformation
\begin{equation}\label{rg12}
\mathcal R:\;\nu\to\nu'
\end{equation}
can be viewed as the RG transformation of the measure $\nu$ in the
Dyson hierarchical model.

If the measure $\nu$ is Lebesgue absolutely continuous, with
a density function $p(s)$, then $\nu'$ is Lebesgue absolutely continuous as well, and
its density function is equal to
\begin{equation}\label{rg13}
\mathcal R(p)(s)=\frac{e^{\be s^2/2^{2-a}}}{Z}\int_{\R} e^{-\be t^2}p\left(\frac{s}{2^{(2-a)/2}}-t\right)
p\left(\frac{s}{2^{(2-a)/2}}+t\right)dt,
\end{equation}
where $Z$ is a normalizing constant,
\begin{equation}\label{rg14}
Z=\int_{\R}e^{\be s^2/2^{2-a}}\int_{\R} e^{-\be t^2}p\left(\frac{s}{2^{(2-a)/2}}-t\right)
p\left(\frac{s}{2^{(2-a)/2}}+t\right)dt\,ds.
\end{equation}

\section{Renormalization group and critical phenomena in the Dyson hierarchical model}

For what follows, it will be convenient to eliminate $\be$ from equation (\ref{rg13}).
To that end we introduce the scaled spin variables,
\begin{equation}\label{sc1}
\tilde\sg(x)=\sqrt{\be}\,\sg(x).
\end{equation}
Then the Gibbs distribution (\ref{v12}) is written as 
\begin{equation}\label{sc2}
d\mu_n(\tilde\sg)=\frac{1}{Z_n}\,e^{-H_n(\tilde\sg)}\prod_{x\in V_n} d\nu\left(\frac{\tilde\sg(x)}{\sqrt{\be}}\right).
\end{equation}
Thus  we eliminate $\be$ in the factor $e^{-H_n(\tilde\sg)}$, and the dependence on $\be$ is included
in the measure $\nu$. In the study of critical phenomena, we consider $\nu$ depending on 
various thermodynamic parameters, like external magnetic field and coupling constants,
 and in this setting $\be$ is one of the thermodynamic parameters. For the sake of brevity,
we denote $\tilde\sg$ in (\ref{sc2}) again by $\sg$, so that    
\begin{equation}\label{sc3}
d\mu_n(\sg)=\frac{1}{Z_n}\,e^{-H_n(\sg)}\prod_{x\in V_n} d\nu(\sg(x)),
\end{equation}
which corresponds to $\be=1$. Respectively, 
the RG transformation of the measure $\nu$ in the
Dyson hierarchical model is given by equation (\ref{rg13}) with $\be=1$, so that
\begin{equation}\label{sc4}
\mathcal R(p)(s)=\frac{e^{s^2/2^{2-a}}}{Z}\int_{\R} e^{- t^2}p\left(\frac{s}{2^{(2-a)/2}}-t\right)
p\left(\frac{s}{2^{(2-a)/2}}+t\right)dt,
\end{equation}
where $Z$ is a normalizing constant,
\begin{equation}\label{sc5}
Z=\int_{\R}e^{s^2/2^{2-a}}\int_{\R} e^{-t^2}p\left(\frac{s}{2^{(2-a)/2}}-t\right)
p\left(\frac{s}{2^{(2-a)/2}}+t\right)dt\,ds.
\end{equation}
We will consider the iterations of the RG transformation,  
\begin{equation}\label{sc6}
p_{m+1}(s)=\mathcal R p_m(s),\qquad m=0,1,\ldots,
\end{equation}
with the initial condition
\begin{equation}\label{sc7}
p_0(s)=\frac{d\nu(s)}{ds}\,.
\end{equation}

In the RG analysis of the critical phenomena, we begin with the search of fixed points of the RG
transformation. First we are looking for Gaussian fixed points.
In this way we find that the Gaussian distribution,
\begin{equation}\label{sc8}
p^*_0(s)=\frac{1}{\sqrt{2\pi \sg}}\,e^{-\frac{s^2}{2\sg}},
\end{equation}
where
\begin{equation}\label{sc9}
\sg=1-2^{1-a},
\end{equation}
is a fixed point of RG transformation (\ref{sc4}). 

To evaluate the stability properties of the Gaussian fixed point $p^*_0(s)$, 
consider the integral operator
\begin{equation}\label{lrg1}
\mathcal L_0 p(s)=\frac{2e^{\be s^2/2^{2-a}}}{Z^*_0}\int_{\R} e^{-\be t^2}p^*_0\left(\frac{s}{2^{(2-a)/2}}-t\right)
p\left(\frac{s}{2^{(2-a)/2}}+t\right)dt,
\end{equation}
where $Z^*_0$ is a normalizing constant,
\begin{equation}\label{lrg2}
Z^*_0=\int_{\R}e^{\be s^2/2^{2-a}}\int_{\R} e^{-\be t^2}p^*_0\left(\frac{s}{2^{(2-a)/2}}-t\right)
p^*_0\left(\frac{s}{2^{(2-a)/2}}+t\right)dt\,ds.
\end{equation}
The operator $\mathcal L_0$ is a linearized RG transformation (\ref{sc4}), in which we set $Z=Z^*_0$.
It is not difficult to
calculate that the eigenvalues of $\mathcal L_0$
in the space of even functions in $L^2(\R)$ are equal to 
\begin{equation}\label{sc10}
\la_j=2^{1-(2-a)j},\qquad j=0,1,2,\ldots,
\end{equation}
and the corresponding eigenfunctions are 
\begin{equation}\label{sc10a}
e_j(s)=G_{2j}(s)p^*_0(s),\qquad j=0,1,2,\ldots,
\end{equation}
where 
\begin{equation}\label{sc10b}
G_{2j}(s)=H_{2j}(\ga s),\qquad \ga=\sqrt{1-2^{a-2}}\,,
\end{equation}
and $H_{2j}(s)$ is the Hermite polynomial of degree $2j$. The first eigenvalue $\la_0=2$
is not essential for the stability property of the fixed point $p^*_0(s)$
with respect to the RG transformation $\mathcal R$,
because it is cancelled by the normalization of $\mathcal R(p)(s)$ in (\ref{sc4}).
Consider the subsequent eigenvalues $\la_1,\;\la_2,\ldots$.

The relevant interval for the parameter $a$ is
\begin{equation}\label{sc11}
1<a<2\,,
\end{equation}
because for $a\le 1$ the free energy is infinite, while for $a\ge 2$ the model does not
exhibits a phase transition, see \cite{Dys1}.

Consider first the parameter $a$ in the interval
\begin{equation}\label{sc12}
1<a<\frac{3}{2}\,.
\end{equation}
Then 
\begin{equation}\label{sc13}
\la_1=2^{a-1}>1;\qquad 1>2^{2a-3}=\la_2>\la_3>\ldots>0,
\end{equation}
hence $p^*_0(s)$ is a fixed point of the hyperbolic type, with one unstable eigenvalue.
Heuristically, one can imagine that there is a one-dimensional unstable invariant manifold 
of the RG transformation, passing through
the fixed point in the space of probability measures, and, in addition, there is a stable invariant manifold of
codimension 1. Therefore, if there is  a one-parameter family of probability measures $\nu_t$, 
which transversally crosses the stable invariant manifold at some value of the parameter, $t=t_c$, then 
the RG iterations of the probability measure $\nu_{t_c}$ should converge
to the Gaussian fixed point, while for $t\not=t_c$, the iterations should go away from the fixed point.
Mathematically, the problem is rather difficult, and it is studied in detail in the works
\cite{BS1}, \cite{B1}. 

To formulate the main result of \cite{BS1}, consider a one-parameter
family of absolutely continuous probability measures 
\begin{equation}\label{sc14}
\{\;d\nu_t(s)=p_t(s)ds,\quad t_1\le t\le t_2\;\},
\end{equation}
such that
for some numbers $C_0,\ldots,C_7>0$, the density $p_t(s)$ satisfies the following conditions, 
when $t\in [t_1,t_2]$ and $\ep=t_2-t_1>0$:
\begin{enumerate}
\item For all $-\infty<s<\infty$,
\begin{equation}\label{sc14a}
p_t(-s)=p_t(s).
\end{equation}
\item On the interval $s\in [-C_0\ep^{-1},C_0\ep^{-1}]$,
\begin{equation}\label{sc14}
p_t(s)=Cp^*_0(s)e^{b_2(t)G_2(s)+b_4(t)G_4(s)+r(s;t)},
\end{equation}
where $G_2(s)$ and $G_4(s)$ are defined in (\ref{sc10b}), and 
$b_2(t),\; b_4(t)$, and $r(s;t)$ satisfy the estimates,
\begin{equation}\label{sc15}
\begin{aligned}
&b_2(t_1)\le -C_1\ep,\quad b_2(t_2)\ge C_2\ep;\qquad b_2'(t)\ge C_3,\\
&-C_4\ep^2\le b_4(t) \le -C_5\ep^2,\quad t\in[t_1,t_2],\\
&|r(s;t)|+\left|\frac{\partial r(s;t)}{\partial s}\right|+\left|\frac{\partial r(s;t)}{\partial t}\right|\le C_6\ep^3, 
\end{aligned}
\end{equation}
for all $t\in[t_1,t_2]$ and $s\in [-C_0\ep^{-1},C_0\ep^{-1}]$. The constant $C=C(t)>0$ is a normalizing factor
to secure that $p_t(s)$ is a probability density.
\item Outside of the interval $[-C_0\ep^{-1},C_0\ep^{-1}]$, the function $p_t(s)$ satisfies
the estimates
\begin{equation}\label{sc16}
p_t(s)+\left|\frac{\partial p_t(s)}{\partial s}\right|+\left|\frac{\partial p_t(s)}{\partial t}\right|
\le Cp^*_0(s)\exp\left(-C_7\ep^2 s^4\right).
\end{equation}
\end{enumerate}

Then the following theorem holds (see \cite{BS1}):

\begin{theo} \label{BS_th1}  Assume that $1<a<\frac{3}{2}\,$ and some numbers $C_0,\ldots,C_7>0$ are given. 
Then there exists $\ep_0>0$ such that if  for some $0<\ep\le\ep_0$, a one-parameter
family of probability densities $p_t(s)$ satisfies conditions (\ref{sc14a})--(\ref{sc16}) with
the constants $C_0,\ldots,C_7$ and some $t_1,\;t_2$, where $t_2-t_1=\ep$,
then there exists a critical point $t_1<t_c<t_2$ such that
\begin{equation}\label{sc17}
\lim_{m\to\infty} \mathcal R^m (p_{t_c})(s)=p^*_0(s).
\end{equation}
\end{theo}

The proof of Theorem \ref{BS_th1} is by induction. In the proof it is shown that there exists a sequence of
intervals $[t_1^{(m)},t_2^{(m)}],\; m=1,2,\ldots$ such that 
\begin{equation}\label{sc18}
t_1=t_1^{(1)}<t_1^{(2)}<\ldots; \qquad t_2=t_2^{(1)}>t_2^{(2)}>\ldots; \qquad
\lim_{m\to\infty} (t_2^{(m)}-t_1^{(m)})=0,
\end{equation}
and the one-parameter
family of probability densities $\mathcal R^m(p_t)(s)$, where $t\in[t_1^{(m)},t_2^{(m)}]$, satisfies
conditions (\ref{sc14})--(\ref{sc16}) with
the constants $C_0,\ldots,C_7>0$ and $\ep=\ep^{(m)}\equiv t_2^{(m)}-t_1^{(m)}$. The critical point is then
\begin{equation}\label{sc19}
t_c=\lim_{m\to\infty} t_1^{(m)}=\lim_{m\to\infty} t_2^{(m)}.
\end{equation}

Theorem \ref{BS_th1} implies that at $t=t_c$ the density of the distribution of the random variable 
\begin{equation}\label{sc20}
\frac{1}{|V_n|^{a/2}}\sum_{x\in V_n}\sg(x)
\end{equation} 
converges to the Gaussian density $p^*_0(s)$. Here $|V_n|=2^n$.
Since $a>1$, this corresponds to a bigger, than in the
central limit theorem, normalization of the sum of the random variables $\sg(x)$. This bigger, nonstandard
normalization is due to long correlations between the random variables $\sg(x)$, and not to
a long tail of the distribution $\nu$. 
The following two theorems proved in \cite{BS1}, \cite{B1}, show that
the critical point is a point of a phase transition from the zero magnetization to a positive magnetization.
Let 
\begin{equation}\label{sc21}
\mathcal G(s;\tau )=\frac{1}{\sqrt{2\pi \tau }}\,e^{-\frac{s^2}{2\tau }}
\end{equation}
be a Gaussian density with the variance $\tau >0$, and let $\mu_n(\sg;\nu)$ be the Gibbs distribution
(\ref{sc3}). 

\begin{theo} \label{BS_th2} Under the assumptions of Theorem \ref{BS_th1},
there exists $\de>0$ such that for any $t_c+\de>t>t_c$,
the density of the distribution of the random variable 
\begin{equation}\label{sc22}
\frac{1}{|V_n|^{1/2}}\sum_{x\in V_n}\sg(x)
\end{equation}
with respect to the Gibbs distribution $\mu_n(\sg;\nu_t)$
converges as $n\to\infty$ to a Gaussian density $\mathcal G(s;\tau (t))$, where the variance $\tau (t)$ behaves like
\begin{equation}\label{sc23}
\tau (t)=\frac{c_0}{t-t_c}\,(1+o(1)),\qquad c_0>0,
\end{equation}
as $t\to t_c+0$.
\end{theo}

\begin{theo} \label{BS_th3} Under the assumptions of Theorem \ref{BS_th1},
there exists $\de>0$ such that for any $t_c>t>t_c-\de$,
there exists a sequence $M_n(t)$ such that
\begin{equation}\label{sc24}
\lim_{n\to\infty} M_n(t)=M(t)>0
\end{equation}
and the densities of the distribution of the random variables 
\begin{equation}\label{sc25}
\frac{1}{|V_n|^{1/2}}\sum_{x\in V_n}\big(\sg(x)+ M_n(t)\big)\quad\textrm{and}\quad 
\frac{1}{|V_n|^{1/2}}\sum_{x\in V_n}\big(\sg(x)- M_n(t)\big)
\end{equation}
with respect to $\mu_n(\sg;\nu_t)$
converge as $n\to\infty$ to a one-half Gaussian density $\frac{1}{2}\,\mathcal G(s;\tau (t)),\;\tau(t)>0$. As $t\to t_c-0$, 
\begin{equation}\label{sc26}
M(t)=c_1(t_c-t)^{1/2}\,(1+o(1)),\qquad \tau (t)=\frac{c_2}{t_c-t}\,(1+o(1)),\qquad c_1,c_2>0.
\end{equation}
\end{theo}

The latter theorem implies the existence of the spontaneous magnetization $M(t)>0$ for $t_c>t>t_c-\de$, so that
\begin{equation}\label{sc27}
\lim_{n\to\infty}\int_{\Sigma_n} \left(\frac{\sum_{x\in V_n}\sg(x)}{|V_n|}\right)^2 d\mu_n(\sg)=M^2(t).
\end{equation}
As $t\to t_c-0$, $M(t)$ has a square-root singularity, which corresponds to the classical Landau theory
of ferromagnetism.

\section{Non-Gaussian fixed points of the renormalization group and non-classical
critical phenomena in the Dyson hierarchical model}

When $a>\frac{3}{2}$, the Gaussian fixed point $p^*_0(s)$ has two or more unstable eigenvalues $\la_j$, $j\ge 1$,
because
\begin{equation}\label{ng1}
\la_1=2^{a-1}>1,\qquad \la_2=2^{2a-3}>1,
\end{equation}
and we are looking for a new, non-Gaussian fixed point $p^*_1(s)$ to describe the critical phenomena at $t=t_c$.
Let 
\begin{equation}\label{ng2}
a=\frac{3}{2}+\ep\,,
\end{equation}
where $\ep>0$ is a small parameter.
We are looking for an RG fixed point as
\begin{equation}\label{ng3}
p^*_1(s)=C p^*_0(s)e^{-c\ep G_4(s)+r(s)}\,,
\end{equation}
where $C>0$ is a normalizing constant, $c>0$ is a constant which can be found from the RG fixed point equation in
the leading order with respect to $\ep$,
and $r(s)$ is a higher order correction, so that on any compact set $K\subset \R$, as $\ep\to 0$,
\begin{equation}\label{ng4}
r(s)=\mathcal O(\ep^2),\qquad s\in K,
\end{equation}
A rigorous proof of the existence of the non-Gaussian fixed point $p^*_1(s)$ for small values of $\ep>0$
is technically quite involved, and it is given in the paper \cite{BS2} of Bleher and Sinai. 
The following result is proven in \cite{BS2}: 

\begin{theo} There exists $\ep_0>0$ such that for any $0<\ep\le \ep_0$, there exists an 
RG fixed point $p_1^*(s)=p_1^*(s;\ep)$ of form \eqref{ng3}, with $r(s)$ satisfying (\ref{ng4})
on any compact set $K\subset \R$. In addition, there exist numbers $C_0,c_0>0$ such that for any $0<\ep<\ep_0$
the function $p_1^*(s)$ satisfies the estimate,
\begin{equation}\label{ng5}
p^*_1(s)\le C_0 p^*_0(s) e^{-c_0\ep s^4}\,,
\end{equation}
\end{theo}

To evaluate the critical exponents associated with the fixed point $p^*_1(s)$, consider the integral
operator, which is a linearized version of the
RG transformation at $p^*_1(s)$:
\begin{equation}\label{ng6}
\mathcal L_1 p(s)=\frac{2e^{\be s^2/2^{2-a}}}{Z^*}\int_{\R^r} e^{-\be t^2}p^*_1\left(\frac{s}{2^{(2-a)/2}}-t\right)
p\left(\frac{s}{2^{(2-a)/2}}+t\right)dt,
\end{equation}
where $Z^*$ is a normalizing constant,
\begin{equation}\label{ng7}
Z^*=\int_{R^r}e^{\be s^2/2^{2-a}}\int_{\R^r} e^{-\be t^2}p^*_1\left(\frac{s}{2^{(2-a)/2}}-t\right)
p^*_1\left(\frac{s}{2^{(2-a)/2}}+t\right)dt\,ds.
\end{equation}
Then we have the following proposition.

\begin{prop} \label{L1} As $\ep\to 0$, the largest three eigenvalues of the operator $\mathcal L_1$ in the
space of even functions in $L^2(\R)$ are $\la_0=2$, $\la_1=\sqrt 2 +O(\ep)$, $\la_2=1-c_0\ep+O(\ep^2)$, where $c_0>0$.
\end{prop}
    
To study the critical phenomena associated with the fixed point $p^*_1(s)$, consider a one-parameter
family of absolutely continuous probability measures 
\begin{equation}\label{ng8}
\{\;d\nu_t(s)=p_t(s)ds,\quad t_1\le t\le t_2\;\},
\end{equation}
such that
for some numbers $C_0,\ldots,C_7>0$, the density $p_t(s)$ satisfies the following conditions, 
when $t\in [t_1,t_2]$ and $t_2-t_1=\ep^{3/2}$:
\begin{enumerate}
\item For all $-\infty<s<\infty$,
\begin{equation}\label{ng9}
p_t(-s)=p_t(s).
\end{equation}
\item On the interval $s\in [-C_0\ep^{-1},C_0\ep^{-1}]$,
\begin{equation}\label{ng10}
p_t(s)=Cp^*_1(s)e^{b_2(t)G_2(s)+b_4(t)G_4(s)+r(s;t)},
\end{equation}
where $G_2(s)$ and $G_4(s)$ are defined in (\ref{sc10b}), and 
$b_2(t),\; b_4(t)$, and $r(s;t)$ satisfy the estimates,
\begin{equation}\label{ng11}
\begin{aligned}
&b_2(t_1)\le -C_1\ep^{3/2},\quad b_2(t_2)\ge C_2\ep^{3/2};\qquad b_2'(t)\ge C_3,\\
&-C_4\ep^2\le b_4(t) \le -C_5\ep^2,\quad t\in[t_1,t_2],\\
&|r(s;t)|+\left|\frac{\partial r(s;t)}{\partial s}\right|+\left|\frac{\partial r(s;t)}{\partial t}\right|\le C_6\ep^3, 
\end{aligned}
\end{equation}
for all $t\in[t_1,t_2]$ and $s\in [-C_0\ep^{-1},C_0\ep^{-1}]$. The constant $C=C(t)>0$ is a normalizing factor
to secure that $p_t(s)$ is a probability density.
\item Outside of the interval $[-C_0\ep^{-1},C_0\ep^{-1}]$, the function $p_t(s)$ satisfies
the estimates
\begin{equation}\label{ng12}
p_t(s)+\left|\frac{\partial p_t(s)}{\partial s}\right|+\left|\frac{\partial p_t(s)}{\partial t}\right|
\le Cp^*_1(s)\exp\left(-C_7\ep^2 s^4\right).
\end{equation}
\end{enumerate}
Then the following theorem holds, see \cite{BS2}.

\begin{theo} \label{BS_th1_NG} Suppose that $a=\frac{3}{2}+\ep\,$, where $0<\ep\le \ep_0$,
 and some numbers $C_0,\ldots,C_7>0$ are given. 
Then there exists $\ep_0>\ep_1>0$ such that if  for some $0<\ep\le\ep_1$, a one-parameter
family of probability densities $p_t(s)$ satisfies conditions (\ref{ng8})--(\ref{ng12}) with
the constants $C_0,\ldots,C_7$ and some $t_1,\;t_2$, where $t_2-t_1=\ep^{3/2}$,
then there exists a critical point $t_1<t_c<t_2$ such that
\begin{equation}\label{ng13}
\lim_{m\to\infty} \mathcal R^m (p_{t_c})(s)=p^*_1(s).
\end{equation}
\end{theo}

Theorem \ref{BS_th1_NG} implies that when $a=\frac{3}{2}+\ep\,$, where $0<\ep\le \ep_1$,
the density of the distribution of the random variable 
\begin{equation}\label{ng13a}
\frac{1}{|V_n|^{a/2}}\sum_{x\in V_n}\sg(x)
\end{equation} 
with respect to the Gibbs distribution $\mu_n(\sg;t_c)$ converges to the non-Gaussian density $p^*_1(s)$.
The next theorem describes the asymptotic behavior of the density of the distribution of
the normalized total spin in the set $V_n$ as $n\to \infty$, when $t>t_c$:

\begin{theo} \label{BS_th2_NG} \cite {BS2} Under the assumptions of Theorem \ref{BS_th1_NG},
if $0<\ep\le\ep_1$, then
there exists $\de>0$ such that for any $t_c+\de>t>t_c$,
the density of the distribution of the random variable 
\begin{equation}\label{ng14}
\frac{1}{|V_n|^{1/2}}\sum_{x\in V_n}\sg(x)
\end{equation}
converges as $n\to\infty$ to a Gaussian density $\mathcal G(s;\tau (t))$. As $t\to t_c+0$,
 the variance $\tau (t)$ behaves like
\begin{equation}\label{ng15}
\tau (t)=\frac{c_0}{(t-t_c)^\ga}\,(1+o(1)),\qquad c_0>0,
\end{equation}
where the critical exponent $\ga$ is equal to
\begin{equation}\label{ng16}
\ga=\frac{a-1}{\log\la_1}\,.
\end{equation}
Here $\la_1>1$ is the eigenvalue of the integral operator $\mathcal L_1$, see Proposition \ref{L1}. 
\end{theo}

Thus for $t_c+\de>t>t_c$ the central limit theorem is valid, with the variance diverging as $t\to t_c$.
The asymptotic behavior of the variance is described by equation \eqref{ng15} with a non-classical critical exponent $\ga$.
The next result concerns with the case $t<t_c$:  

\begin{theo} \label{BS_th3_NG} \cite{BS2} Under the assumptions of Theorem \ref{BS_th1_NG},
if $0<\ep\le\ep_1$, then
there exists $\de>0$ such that for any $t_c>t>t_c-\de$,
there exists a sequence $M_n(t)$ such that
\begin{equation}\label{ng17}
\lim_{n\to\infty} M_n(t)=M(t)>0
\end{equation}
and the densities of the distribution of the random variables 
\begin{equation}\label{ng18}
\frac{1}{|V_n|^{1/2}}\sum_{x\in V_n}\big(\sg(x)+ M_n(t)\big)\quad\textrm{and}\quad 
\frac{1}{|V_n|^{1/2}}\sum_{x\in V_n}\big(\sg(x)- M_n(t)\big)
\end{equation}
converge as $n\to\infty$ to a one-half Gaussian density $\frac{1}{2}\,\mathcal G(s;\tau (t)),\;\tau(t)>0$. As $t\to t_c-0$, 
\begin{equation}\label{ng19}
M(t)=c_1(t_c-t)^{\be}\,(1+o(1)),\qquad \tau (t)=\frac{c_2}{(t_c-t)^\ga}\,(1+o(1)),\qquad c_1,c_2>0,
\end{equation}
where the critical exponents $\be$ and  $\ga$ are equal to
\begin{equation}\label{ng20}
\be=\frac{2-a}{2\log\la_1}\,,\qquad            \ga=\frac{a-1}{\log\la_1}\,,
\end{equation}
where $\la_1$ is the eigenvalue of the integral operator $\mathcal L_1$.
\end{theo}

Again, the critical exponents $\be$ and $\ga$ given by equation \eqref{ng20} take non-classical values.

\section{Conclusion}

The Dyson hierarchical model is of a great interest because for this model the RG group transformation reduces to 
nonlinear integral equation \eqref{rg13}, and this allows a study of critical phenomena unavailable in other
models. The results of the works \cite{BS1}, \cite{BS2}, \cite{B1} reviewed in this paper, give a rigorous proof 
for the general, mostly heuristic principles of the renormalization group method in the theory of
critical phenomena, developed in physical works of Wilson, Fisher, Kadanoff, and others 
(see the works \cite{WK}, \cite{Fis} and
references therein). Since then, the works \cite{BS1}, \cite{BS2}, \cite{B1} has been extended in several directions. 
Let us briefly mention some of these extensions. 

A new proof of the convergence of the RG transformations for $a<\frac{3}{2}$ is given in the monograph \cite{Sin}
of Sinai.

The marginal case $a=\frac{3}{2}$ is studied in the work \cite{B2}. It is shown in \cite{B2}, that
if $a=\frac{3}{2}$ then 
at the critical point $t_c$, the RG iterations $\mathcal R^n p_{t_c}(s)$  converge
to the Gaussian fixed point $p^*_0(s)$, but the convergence is slow, polynomial in $n$. In addition, the critical exponents
$\be$ and $\ga$ take the classical values $\be=\frac{1}{2}$ and $\ga=1$, but there are logarithmic corrections
in the asymptotic behavior of the spontaneous magnetization and the variance as $t\to t_c$. 
These logarithmic corrections are similar to the ones appearing in quantum field theory in
critical dimension. 

Strong large deviation asymptotics for the average spin in the Dyson hierarchical model
are derived in the work \cite{B3}. It is shown in \cite{B3} that
the large deviation function $\Phi(s;t)$ is strictly convex for $t>t_c$, with a positive curvature
at all $s\in \R$. At $t=t_c$, it is still convex but the curvature is equal to 0 at $s=0$.
Finally, for $t<t_c$, $\Phi(s;t)$ is strictly convex for $|s|\ge M(t)$,
where $M(t)>0$ is the spontaneous magnetization, and it is constant on the interval $-M(t)\le s\le M(t)$.
This is the Maxwell phenomenon in the theory of phase transitions. The surface tension corrections
to the canonical free energy in the coexistence interval  $[-M(t),M(t)]$ are evaluated in the work
\cite{B4}.

The results  of the works \cite{BS1}, \cite{BS2}, \cite{B1} are extended to the Dyson vector-valued
hierarchical model. In this case the spin variables $\sg(x)$ take value in the space $\R^d$, $d\ge 2$, 
and the measure $\nu$ is assumed to be invariant with respect to orthogonal transformations in $\R^d$.
A new important phenomenon, related to the continuous symmetry of the orthogonal group $O(d)$, arises
for $d\ge 2$. It is a slow, polynomial decay of correlations at $t<t_c$, which leads to the convergence
of the RG transformations with a non-standard normalization to a fixed point at $t<t_c$. This phenomenon is related 
to the Goldstone modes in the theory of phase transitions with continuous symmetry. A detailed
study of the  RG fixed points and the RG convergence to these fixed points at $t<t_c$ is given
in the series of works of Bleher and Major \cite{BM1}--\cite{BM4}.


\begin{thebibliography}{9999}




\bibitem{B1} 
P. Bleher, Second order phase transition in some models of ferromagnetism. {\it Trans.
          Moscow Math. Soc.} {\bf 33} (1975), 155--222. 

\bibitem{B2} 
P. Bleher, On second order phase transitions in Dyson's asymptotically hierarchical models.
{\it Uspekhi Math. Nauk} {\bf 32} (1977), 243--244.

\bibitem{B3} 
P. Bleher, Large deviations theorem in the vicinity of the critical point of the 
$\f^4$-hierarchical model. Theor. Probab.  Its Appl.  {\bf 30} (1985), 499--510. 

\bibitem{B4} 
P. Bleher, Phase separation phenomenon in the $\f^4$-hierarchical model. Theor.  Math.
          Phys. {\bf 11} (1984), 226--240. 
 
\bibitem{BM1} 
P. Bleher and P. Major, Renormalization of Dyson's
hierarchical vector valued $\f^4$-model at low temperatures. {\it
Commun. Math. Physics\/} {\bf 95}  (1984), 487--532.
 
\bibitem{BM2} 
P. Bleher and P. Major,  Critical phenomena and
universal exponents in statistical physics. On Dyson's hierarchical
model. {\it Ann. Probab.\/} {\bf15} (1987), 431--477.
 
\bibitem{BM3} 
P. Bleher and P. Major,  The large-scale limit of
Dyson's hierarchical vector-valued model at low temperatures. The
non-Gaussian case. Parts I, II. {\it Annales de l'Institut Henri
Poincar\'e,} S\'erie Physique Th\'eorique, Volume
{\bf49} (1988) 1--85,  87--143.


\bibitem{BM4}  P. Bleher and P. Major,  The large-scale limit of
Dyson's hierarchical vector-valued model at low temperatures. The
marginal case $c=\sqrt2$. {\it Comm. Math. Physics\/} {\bf 125} (1989), 43--69.

\bibitem{BS1} 
P. Bleher and Ya.G. Sinai, Investigation of the critical point in models of the type of
          Dyson's hierarchical model. {\it Commun. Math. Phys.} {\bf 33} (1973), 23--42. 

\bibitem{BS2} 
P. Bleher and Ya.G. Sinai, Critical exponents for Dyson's asymptotically hierarchical models. 
{\it Commun. Math. Phys.} {\bf 45} (1975),  247--278. 
 
\bibitem {Dys1} F.J. Dyson,
Existence of a phase transition in a one-dimensional Ising
ferromagnet. {\it Commun. Math. Phys.}\/ {\bf 12} (1969), 91--107.
 
\bibitem {Dys2} F.J. Dyson,
An Ising ferromagnet with discontinuous long-range order.
{\it Commun. Math. Phys.}\/ {\bf 21} (1971), 269--283.
 
\bibitem {Dys3} F.J. Dyson,
Existence and nature of phase transitions in one-dimensional Ising
ferromagnets. {\it Mathematical aspects of statistical mechanics
(Proc. Sympos. Appl. Math., New York, 1971),} pp. 1--12, SIAM-AMS
Proceedings, Vol. V, {\it Amer. Math. Soc., Providence, R.I.,} (1972).
 
\bibitem {Fis} M.E. Fisher, The theory of equilibrium critical phenomena.
{\it Rep. Progr. Phys.} {\bf 30} (1967), 615--730.

\bibitem{Sin}
Ya.G. Sinai, Theory of phase transitions,
Theory of phase transitions: rigorous results. 
International Series in Natural Philosophy, 108. Pergamon Press, Oxford-Elmsford, N.Y., 1982. viii+150 pp.
 
\bibitem {WK} K.G. Wilson and J. Kogut:
The renormalization group and the $\ep$-expansion.
{\it Phys. Rep.} {\bf 12C} (1974), 75--199.
 
 
 
 

\end{thebibliography}
\end{document}